\documentclass[11pt,letterpaper]{article}

\usepackage[T1]{fontenc}
\usepackage[utf8]{inputenc}
\usepackage{mathpazo}                   
\usepackage[scaled=0.90]{helvet}        
\usepackage{courier}                    

\usepackage{amsmath, amssymb, amsthm}
\usepackage{booktabs}
\usepackage{tabularx}
\usepackage{graphicx}
\usepackage{microtype}
\usepackage{hyperref}
\usepackage{xcolor}
\usepackage{geometry}
\usepackage{enumitem}
\usepackage{csquotes}

\hypersetup{
    colorlinks=true,
    linkcolor=darkgray,
    citecolor=darkgray,
    urlcolor=darkgray
}

\newtheorem{definition}{Definition}

\newtheorem{observation}{Observation}

\title{\textbf{What Distributed Computing Got Wrong:} \\
The Category Mistake That Turned Design Choices \\
into Laws of Nature}

\author{Paul Borrill \\ D{\AE}D{\AE}LUS}
\date{02026-FEB-20 \\ \small v\,0.01}

\begin{document}

\maketitle

\begin{abstract}
The foundational impossibility results of distributed computing---the
Fischer-Lynch-Paterson theorem, the Two Generals Problem, the CAP
theorem---are widely understood as discoveries about the physical limits
of coordination.  This paper argues that they are nothing of the sort.
They are consequences of a \emph{category mistake}: treating
Forward-In-Time-Only (FITO) information flow as a law of nature rather
than recognizing it as a design choice inherited from Shannon's channel
model and Lamport's happened-before relation.  We develop this argument
in six steps.  First, we introduce the category mistake framework from
Ryle through Spekkens' ontic/epistemic distinction in quantum
foundations.  Second, we identify FITO as the hidden axiom that unifies
the classical impossibility results.  Third, we apply Spekkens'
Leibnizian principle to show that FITO-based models contain surplus
ontological structure.  Fourth, we develop the counterfactual: what
changes when FITO is dropped.  Fifth, we demonstrate that the
impossibility theorems are theorems about FITO systems, not about
physics.  Sixth, we sketch the transactional alternative---bilateral
interactions that dissolve the apparent impossibilities by replacing
unidirectional message passing with atomic bilateral transactions.  The
implication is that distributed computing has spent fifty years
optimizing within the wrong design space.
\end{abstract}

\section{Introduction}
\label{sec:intro}

Distributed computing has a canon.  At its centre sit three results
that every graduate student learns as fundamental limits on what
distributed systems can achieve:

\begin{enumerate}[label=(\roman*)]
\item \textbf{FLP} (Fischer, Lynch, Paterson, 1985): No deterministic
  protocol can guarantee consensus in an asynchronous system with even
  one faulty process~\cite{flp1985}.
\item \textbf{Two Generals} (formalized by Halpern and Moses, 1990):
  Common knowledge is unattainable in asynchronous message-passing
  systems~\cite{halpernmoses1990}.
\item \textbf{CAP} (Brewer, 2000; Gilbert and Lynch, 2002): A
  distributed system cannot simultaneously provide consistency,
  availability, and partition tolerance~\cite{brewer2000cap,gilbert2002cap}.
\end{enumerate}

These results are treated as discoveries about the physical world---as
constraints imposed by nature on any conceivable distributed system, no
matter how cleverly engineered.  Textbooks present them alongside the
second law of thermodynamics and the speed of light: hard limits that
cannot be circumvented, only accommodated.

This paper argues that they are not hard limits.  They are consequences
of a modelling assumption so deeply embedded in the field that it has
become invisible: the assumption that information flows
\emph{forward in time only} (FITO), from sender to receiver, from cause
to effect, through channels that add noise to signals rather than
interacting with them.

The assumption is not wrong.  It is \emph{limited}.  And the
impossibility results are theorems about FITO systems, not theorems
about physics.  Recognizing this distinction---identifying the category
mistake---opens a design space that the field has not explored.

\subsection{Structure of the Argument}

The argument proceeds in six steps.
Section~\ref{sec:category} introduces the category mistake framework
from Ryle through Spekkens' work in quantum foundations.
Section~\ref{sec:fito} identifies FITO as the hidden axiom unifying the
classical impossibility results.
Section~\ref{sec:leibniz} applies Spekkens' Leibnizian principle to show
that FITO models contain surplus ontological structure.
Section~\ref{sec:counterfactual} develops the counterfactual: what
changes when FITO is treated as a design choice rather than a physical
law.
Section~\ref{sec:dissolve} demonstrates that the impossibility theorems
dissolve under the corrected categorization.
Section~\ref{sec:transactional} sketches the transactional alternative.

\section{The Category Mistake Framework}
\label{sec:category}

\subsection{Ryle: The University That Isn't a Building}

Gilbert Ryle introduced the concept of a category mistake in
\emph{The Concept of Mind} (1949).  His canonical example: a visitor
tours Oxford's colleges, libraries, playing fields, and administrative
offices, then asks, ``But where is the University?''  The visitor has
made a category mistake---treating ``the University'' as if it were an
entity of the same category as the buildings, when it is actually the
\emph{organizational structure} that relates
them~\cite{ryle1949concept}.

Category mistakes are not errors of fact.  The visitor has seen all the
relevant buildings.  The error is about the \emph{type} of thing under
discussion.  Category mistakes generate pseudo-problems: puzzles that
seem profound but dissolve once the categories are corrected.

\subsection{Spekkens: The Ontic/Epistemic Distinction}

Robert Spekkens, at the Perimeter Institute, has developed the most
sustained analysis of category mistakes in modern
physics~\cite{spekkens2018sphinx,harrigan2010einstein}.  His programme
rests on a single distinction: is a given theoretical entity
\emph{ontic} (a direct description of physical reality) or
\emph{epistemic} (a representation of an observer's knowledge about
physical reality)?

In his 2007 toy model, Spekkens demonstrated that an astonishing range
of ``quintessentially quantum'' phenomena---interference, entanglement,
teleportation, no-cloning---can be reproduced in a classical system with
a single epistemic restriction: the \emph{knowledge balance principle},
which states that maximal knowledge always leaves as much unknown as
known~\cite{spekkens2007toy}.

The phenomena that \emph{cannot} be reproduced---Bell inequality
violations and Kochen-Specker contextuality---mark the boundary of
genuine quantum physics, distinguishing it from artefacts of category
confusion.

The lesson is precise: when an epistemic construct (the quantum state
$\psi$) is mistaken for an ontic entity (a complete description of
physical reality), pseudo-problems proliferate---wave function collapse,
the measurement problem, spooky action at a distance.  When the category
is corrected, many puzzles dissolve.

\subsection{From Quantum Foundations to Distributed Systems}

This paper applies Spekkens' programme to distributed computing.  The
claim: just as many quantum ``mysteries'' arise from treating
$\psi$~as~ontic when it is epistemic, many distributed systems
``impossibilities'' arise from treating FITO message passing as a
physical law when it is a design choice.

The analogy is not metaphorical.  It is structural:

\begin{center}
\begin{tabular}{lll}
\toprule
\textbf{Domain} & \textbf{Ontic Reading (Mistake)} & \textbf{Epistemic Reading (Correct)} \\
\midrule
Quantum mechanics & $\psi$ describes reality & $\psi$ describes knowledge \\
Information theory & Message is a physical entity & Message is a representation \\
Lamport clocks & Happened-before is causation & Happened-before is message path \\
Distributed systems & Impossibilities are physical & Impossibilities are model-relative \\
\bottomrule
\end{tabular}
\end{center}

\section{FITO: The Hidden Axiom}
\label{sec:fito}

\subsection{Formal Definition}

Forward-In-Time-Only (FITO) is the assumption that causation, ordering,
and correctness in physical and computational systems can be defined by
a unidirectional progression of time.  It manifests as three interlocking
commitments~\cite{borrill2026fito}:

\begin{definition}[The FITO Triad]
\leavevmode
\begin{enumerate}[label=(\roman*)]
\item \textbf{Temporal Monotonicity.}  Time only advances.  State
  changes are accumulated, never subtracted.  Every FITO system embeds a
  partial order $\prec$ on events such that $a \prec b \implies t(a) <
  t(b)$, where $t(\cdot)$ is a timestamp function.  There is no
  backward edge in the causal graph.

\item \textbf{Asymmetric Causation.}  Causes lie in the temporal past of
  their effects.  A system in state $S_t$ can influence $S_{t+\Delta}$
  but not $S_{t-\Delta}$.  The transition function is a forward map:
  $f: S_t \to S_{t+\Delta}$, $\Delta > 0$.

\item \textbf{Global Temporal Reference.}  ``Earlier'' and ``later'' are
  assumed to be globally meaningful.  There exists, at least in
  principle, a total ordering of events that all observers could agree
  upon if only their clocks were sufficiently synchronized.
\end{enumerate}
\end{definition}

Together, these form the implicit foundation of nearly all engineering
systems---and the explicit foundation of the impossibility proofs in
distributed computing.

\subsection{FITO in Shannon's Channel}

Claude Shannon's 1948 channel model~\cite{shannon1948} encodes FITO at
the deepest level.  The model: Source $\to$ Encoder $\to$ Channel $\to$
Decoder $\to$ Destination.  Information flows unidirectionally, from
sender to receiver, through a channel that adds noise \emph{to} the
signal rather than interacting \emph{with} it:
\[
Y = X + N
\]
where $X$ is the transmitted signal, $N$ is noise (independent of $X$
and the receiver), and $Y$ is the received signal.  The receiver is
passive.  There is no return path in the basic model; feedback, if
present, is a separate construction.

Yet mutual information is perfectly symmetric: $I(X;Y) = I(Y;X)$.  The
information measure sees no directionality.  The asymmetry is not in the
mathematics of information---it is in the \emph{temporal structure
imposed on the model}.

Physical communication is not like this.  Electromagnetic interaction is
bilateral: the receiver's antenna radiates, the sender's antenna
receives.  The channel is a mutual impedance field, not a one-way pipe.
Shannon's model abstracts this away, and every communication protocol
built on Shannon's framework inherits the abstraction.

\subsection{FITO in Lamport's Happened-Before}

Leslie Lamport's 1978 paper~\cite{lamport1978time} established the
theoretical vocabulary of distributed computing.  His happened-before
relation ($\to$) defines potential causality:

\begin{enumerate}[label=(\arabic*)]
\item If $a$ and $b$ are events in the same process and $a$ comes before
  $b$, then $a \to b$.
\item If $a$ is the sending of a message and $b$ is the receipt of that
  message, then $a \to b$.
\item If $a \to b$ and $b \to c$, then $a \to c$ (transitivity).
\end{enumerate}

Rule~(2) is FITO encoded as a partial order: sending precedes receiving.
There is no mechanism for the receiver's state to constrain the
sender's.  The causal arrow is one-way.  Logical clocks are monotonic
counters that track this order.  They cannot decrease.  They cannot
represent ``$b$ influenced $a$ even though $t(b) > t(a)$.''

This is not a limitation Lamport overlooked.  It is a foundational
commitment: the model of communication is unidirectional message
passing, and the model of time is the partial order induced by that
passing.  The commitment is so natural that it appears to be physics.
But it is not physics.  It is a model.

\subsection{FITO in the Impossibility Proofs}

Every major impossibility result in distributed computing is proved
within a model that assumes FITO communication:

\begin{center}
\begin{tabular}{lll}
\toprule
\textbf{Result} & \textbf{Model} & \textbf{FITO Assumption} \\
\midrule
FLP~\cite{flp1985} & Asynchronous message passing & One-way, arbitrarily delayed \\
Two Generals~\cite{halpernmoses1990} & Unreliable message passing & Unidirectional, lossy \\
CAP~\cite{gilbert2002cap} & Network partitions & Messages in flight, no bilateral \\
Herlihy~\cite{herlihy1991waitfree} & Read/write registers & Unilateral reads and writes \\
\bottomrule
\end{tabular}
\end{center}

In each case, the communication primitive is a \emph{message}---an
entity that is created at one endpoint, exists independently ``in
flight,'' and may or may not arrive at another endpoint.  There is no
primitive in which both endpoints participate simultaneously.  There is
no bilateral transaction.

This is the hidden axiom.  It is so pervasive that identifying it
requires the same kind of conceptual archaeology that Spekkens performed
on quantum foundations: examining what everyone assumed and asking
whether it could be otherwise.

\section{The Leibnizian Principle Applied}
\label{sec:leibniz}

\subsection{Spekkens' Methodological Criterion}

In a 2019 paper, Spekkens defended a methodological principle he traces
to Leibniz: \emph{if an ontological theory predicts two scenarios that
are ontologically distinct but empirically indiscernible, then the theory
should be rejected in favour of one where the scenarios are
ontologically identical}~\cite{spekkens2019leibniz}.

This is Leibniz's identity of indiscernibles repurposed as a criterion
for theory construction.  Spekkens argues that Einstein applied this
principle repeatedly: special relativity eliminates the distinction
between ``truly at rest'' and ``moving at constant velocity'' because
the scenarios are empirically indiscernible; general relativity
eliminates the distinction between gravitational and inertial mass for
the same reason.

\subsection{Surplus Structure in Distributed Computing}

The Leibnizian principle applies directly to distributed systems.
Consider two FITO models of a distributed computation:

\begin{enumerate}[label=(\alph*)]
\item Model~A assigns timestamps from a Lamport clock.
\item Model~B assigns timestamps from a vector clock.
\end{enumerate}

Both models predict the same observable behaviour---the same set of
completed transactions, the same final states, the same external
outputs.  They differ only in the internal structure they ascribe to
``time'' in the system.

By Spekkens' principle, the ontological differences between these models
(scalar vs.\ vector timestamps, different causal orderings of concurrent
events) are surplus structure.  They are features of the
\emph{model}, not of the \emph{system}.  Any engineering decision that
depends on which timestamp model is ``correct'' is building on a
model-dependent epistemic artefact.

\subsection{The Deeper Surplus: FITO Itself}

The Leibnizian argument cuts deeper.  Consider two models of the same
distributed system:

\begin{enumerate}[label=(\alph*)]
\item A FITO model where messages travel from sender to receiver, with
  the possibility of loss.
\item A bilateral model where communication is an atomic interaction
  between two endpoints, with the possibility of non-interaction.
\end{enumerate}

If both models predict the same observable outcomes---the same
successful and failed coordination attempts, the same final states---then
they are empirically indiscernible.  The FITO model's
ontological commitment to ``messages in flight''---entities that exist
independently of both sender and receiver---is surplus structure.

The ``message in flight'' is the distributed systems analogue of the
``ether'' in pre-relativistic physics: an ontological commitment that
the theory requires but that no experiment can detect.  Einstein
eliminated the ether by recognizing it as surplus structure.  The same
move is available in distributed computing.

\subsection{Helland's Observation}

Pat Helland has articulated the practical consequence of this surplus
structure, though without using the Leibnizian
vocabulary~\cite{helland2015immutability}.  His aphorism---``Writes are
wrong and reads are worse''---identifies the operational failure mode: a
write is a unilateral declaration of intent with no guarantee of
reception or coordination; a read creates invisible dependencies on
potentially stale state.

Helland's critique is the engineer's version of the Leibnizian argument.
The read/write primitives commit to a model (unilateral state
modification, unilateral state inspection) that carries ontological
baggage no experiment can justify.  Herlihy proved the mathematical
consequence: read/write registers have consensus number~1---they cannot
solve consensus between even two
processes~\cite{herlihy1991waitfree}.  The limitation is not engineering
precision.  It is ontological commitment.

\section{The Counterfactual: What If FITO Fails?}
\label{sec:counterfactual}

The counterfactual is not speculation.  It is the discipline of examining
what changes when FITO is relaxed---when we treat it as a modelling
assumption rather than a physical necessity.

\subsection{In Physics: Time-Symmetric Dynamics}

Every fundamental equation of physics is symmetric under time reversal
(modulo CPT in quantum field theory).  Newton's $F = ma$ is unchanged
under $t \to -t$.  The Schr\"odinger equation has solutions in both
temporal directions.  Maxwell's equations admit advanced as well as
retarded solutions---waves converging toward a source rather than
radiating from it~\cite{wheeler1945absorber}.

The time-asymmetry we observe---entropy increase, radiation from
sources, causes before effects---is not \emph{in} the laws.  It is in
the boundary conditions we impose and the interpretation we overlay.

Bell's theorem~\cite{bell1964epr} proves that no local
hidden-variable theory with FITO causality can reproduce quantum
correlations.  The standard conclusion: quantum mechanics is nonlocal.
But the theorem has a fourth, usually unstated assumption---FITO
causality.  If future measurement settings can influence past hidden
variables (retrocausality), Bell correlations arise without
nonlocality~\cite{price2012retrocausality,leifer2017retrocausality}.
This is not exotic speculation.  It is a logical consequence of taking
the time-symmetry of the fundamental equations seriously.

\subsection{In Distributed Computing: Bilateral Interactions}

The physics offers a template.  Wheeler and Feynman's absorber
theory~\cite{wheeler1945absorber} provides the model: radiation is not
``emitted'' until it is ``absorbed.''  Neither event is meaningful
without the other.  The transaction spans both endpoints as a single
bilateral act.  Cramer's transactional interpretation~\cite{cramer1986transactional}
extends this to all quantum phenomena: offer wave forward, confirmation
wave backward, transaction complete or transaction does not exist.

The distributed systems counterfactual: replace the \emph{message}
(a unilateral, FITO entity) with the \emph{bilateral transaction}
(an atomic interaction that either completes for both endpoints or
fails for both).

This is not a minor variation.  It changes the algebraic structure of
the system.  FITO communication lives in a \emph{semiring}: operations
accumulate but cannot subtract.  Clocks tick up, counters increment,
logs append.  There is no mechanism to ``uncount'' or ``untick.''
Bilateral communication lives in a \emph{group}: operations have
inverses.  Shared channel properties cancel under subtraction.
Synchronization error is eliminated, not merely
averaged~\cite{borrill2026fito}.

\begin{observation}[The Causal Cancellation Principle]
On a bidirectional channel, $d_{AB} - d_{BA}$ eliminates the shared
propagation delay $d_0$.  FITO clocks can only compute
$d_{AB} + d_{BA}$, which doubles it.  The limitation is not engineering
precision---it is algebraic structure.
\end{observation}

\section{The Impossibility Theorems Dissolve}
\label{sec:dissolve}

We do not claim that the impossibility theorems are wrong.  They are
correct---within their models.  What we claim is that they are theorems
about \emph{FITO systems}, not theorems about \emph{physics}.  The
distinction matters because it means the impossibilities are contingent
on a design choice, not imposed by nature.

\subsection{FLP Reconsidered}

The Fischer-Lynch-Paterson proof~\cite{flp1985} proceeds by showing that
from any bivalent configuration (one where both outcomes are still
reachable), an adversary can always extend the execution to maintain
bivalency, preventing agreement.

Examine the model:

\begin{enumerate}[label=(\alph*)]
\item Processes communicate by \textbf{message passing}.
\item Messages may be \textbf{arbitrarily delayed}.
\item There is \textbf{no bound} on relative process speeds.
\end{enumerate}

Every message is a one-way, fire-and-forget transmission.  The sender's
state advances; the message is ``in flight''; the receiver's state may
eventually advance.  There is no bilateral handshake, no atomic
transaction that completes for both parties or neither.

FLP proves that \emph{under this model}, consensus is impossible.  The
impossibility is real, but it is an impossibility within FITO
message-passing.  It does not prove that \emph{coordination itself} is
impossible---only that coordination via one-way messages with unbounded
delay cannot guarantee termination.

In a bilateral model, the adversary's power is different.  The adversary
can prevent a transaction from completing (both parties remain at their
prior state), but it cannot create a state where one party has advanced
and the other has not.  The bivalency argument requires the ability to
``lose'' a message that has already been sent but not yet received---a
state that does not exist in a bilateral model, because there are no
messages ``in flight.''

\subsection{Two Generals Reconsidered}

Halpern and Moses~\cite{halpernmoses1990} formalized the Two Generals
Problem: in asynchronous message-passing systems, common knowledge is
unattainable.  The proof depends on the infinite regress of
acknowledgments: $A$ sends a message to $B$; $B$ acknowledges; $A$
acknowledges the acknowledgment; and so on, ad infinitum.  At each
stage, one party is uncertain whether the most recent message was
received.

The key assumption: communication is unidirectional.  Each message is a
separate act---it may be lost, and the sender cannot know whether it
arrived without receiving a separate acknowledgment, which itself may be
lost.

In a bilateral model, the question transforms.  The question is not
``does $B$ know that $A$ knows?'' but ``did the coordination transaction
complete?''  A completed bilateral transaction \emph{constitutes} common
knowledge of the exchanged information.  An incomplete transaction
leaves both parties at their prior state.  There is no intermediate
condition.

The infinite regress collapses because the transaction is bilateral from
the start.  The Halpern-Moses impossibility is conditional on the
message-passing model.  In a transactional model, the assumption of
unidirectional messages is replaced by bilateral completion, and the
impossibility no longer applies.

\subsection{CAP Reconsidered}

Brewer's CAP theorem~\cite{brewer2000cap} and its formalization by
Gilbert and Lynch~\cite{gilbert2002cap} state that a distributed system
cannot simultaneously guarantee consistency (all nodes see the same
data), availability (every request receives a response), and partition
tolerance (the system continues to operate despite network partitions).

The proof assumes that during a partition, a write at one node cannot
be communicated to another node---the standard FITO message-passing
model.  The system must choose: accept the write on one side of the
partition (sacrificing consistency) or reject it (sacrificing
availability).

In a bilateral model, the question changes.  A ``write'' is not a
unilateral state modification but a bilateral transaction.  During a
partition, the transaction between nodes on opposite sides cannot
complete---which means it does not happen.  Both nodes remain at their
prior state.  There is no inconsistency (both agree on the prior state)
and no lost availability (both can continue to process local bilateral
transactions).

The CAP impossibility arises from the FITO assumption that a write can
occur at one node independently of other nodes---that state
modification is unilateral.  When state modification is bilateral, the
three-way trade-off dissolves into a two-party coordination problem
with a well-defined outcome: complete or don't.

\subsection{Herlihy's Hierarchy Reconsidered}

Herlihy's consensus hierarchy~\cite{herlihy1991waitfree} proves that
read/write registers have consensus number~1: they cannot solve
wait-free consensus between two processes.  The proof exploits the
commutativity of reads and writes across independent memory
locations---if $P_0$ writes $R_0$ and $P_1$ writes $R_1$, the final
state is identical regardless of order, and neither process can detect
which happened first.

But compare-and-swap (CAS) and memory-to-memory swap have
\emph{infinite} consensus number.  What distinguishes them from
reads and writes?

The answer: swap is a bilateral primitive.  A swap atomically
exchanges values between two locations---both locations participate in
a single indivisible operation.  The ``message'' is not created at one
endpoint and consumed at another; the operation spans both endpoints
simultaneously.  CAS is bilateral in a weaker sense: the comparison and
the conditional write happen as a single atomic unit, creating a
connection between the prior state and the update that reads and writes
cannot achieve.

Herlihy's hierarchy is a map of the FITO landscape.  Weak
(unilateral) primitives sit at the bottom; strong (bilateral) primitives
sit at the top.  The hierarchy is real, but it is a hierarchy of
\emph{design primitives}, not of physical possibility.

\section{The Transactional Alternative}
\label{sec:transactional}

If the impossibility results are contingent on FITO, what does the
alternative look like?

\subsection{The Wheeler-Feynman Template}

Wheeler and Feynman's absorber theory~\cite{wheeler1945absorber}
provides the template: a photon is not ``emitted'' until it is
``absorbed.''  Neither event is meaningful without the other.  Cramer's
transactional interpretation~\cite{cramer1986transactional} generalizes:
every quantum event is a completed transaction between emitter and
absorber.  There are no photons ``in flight'' in the ontological sense;
there are only completed transactions.

\subsection{Open Atomic Ethernet}

Open Atomic Ethernet (OAE)~\cite{oae2024spec} applies this insight
to network architecture.  Instead of Shannon's unidirectional
$X \to Y$, OAE implements bilateral $X \leftrightarrow Y$.  Every
frame carries information in both directions simultaneously---the
acknowledgment of previous transmissions is embedded in the current
transmission, not sent on a separate channel.

\begin{description}[style=nextline]
\item[Bilateral atomicity.]
  An OAE transaction either completes for both endpoints or fails for
  both.  There is no intermediate state where data is ``in flight.''
  This is not ``zero duration'' but ``mutual completion''---both
  parties finish together or neither does.

\item[Memory-to-memory swap.]
  The primitive operation is swap, not read/write.  Swap has infinite
  consensus number~\cite{herlihy1991waitfree}.  It can solve consensus
  among any number of processes---not because it is ``faster'' or
  ``more reliable'' than reads and writes, but because it is
  \emph{bilateral}: both memory locations participate in a single
  indivisible act.

\item[Interaction-derived ordering.]
  The temporal order of events is not imposed by timestamps or logical
  clocks.  It emerges from the pattern of completed transactions.
  Two events are ordered if and only if one is part of a completed
  transaction that the other causally depends on.  If no transaction
  connects them, they are genuinely concurrent---not ``concurrently
  ordered by some unknown timestamp'' but structurally independent.
\end{description}

\subsection{Dissolving the Compensation Mechanisms}

FITO systems require unbounded compensation mechanisms to cope with
the consequences of unilateral communication: timeout and retry (TAR),
two-phase commit (2PC), Paxos, Raft, and the entire apparatus of
consensus protocols.  Each mechanism adds complexity, latency, and
failure modes.

In a bilateral model, many of these mechanisms are unnecessary:

\begin{center}
\begin{tabular}{lll}
\toprule
\textbf{FITO Mechanism} & \textbf{Problem Addressed} & \textbf{Bilateral Resolution} \\
\midrule
Timeout and retry & Message loss uncertainty & Transaction completes or doesn't \\
Two-phase commit & Atomic multi-party update & Bilateral swap is already atomic \\
Consensus protocols & Agreement despite faults & Transaction \emph{constitutes} agreement \\
Vector clocks & Causal ordering & Order emerges from transactions \\
CRDTs & Eventual consistency & Bilateral consistency is immediate \\
\bottomrule
\end{tabular}
\end{center}

The compensation mechanisms are not solutions to physical problems.
They are patches for problems created by the FITO assumption.  When the
assumption is relaxed, the problems---and the need for the
patches---diminish.

\subsection{Pratt's Pomsets: Partial Orders as Partial Solution}

Vaughan Pratt's pomsets~\cite{pratt1986pomsets}---partially ordered
multisets as semantic objects for concurrency---represent an important
step away from interleaving semantics.  Pomsets separate causal
constraint ($e \le f$: ``$e$ must happen before $f$'') from
time-of-observation (when an observer sees events).  This distinction
is essential.

But pomsets embed FITO through the antisymmetry of their partial order:
if $e \le f$ and $f \le e$, then $e = f$.  Causal loops are
forbidden by construction.  This makes pomsets a necessary but
insufficient foundation.  They escape the interleaving fallacy but
remain within the FITO design space.

The transactional alternative requires richer structures: event
structures with conflict and enabling
relations~\cite{winskel1987events}, symmetric process calculi that
model forward and backward transitions, and conserved-quantity
frameworks that embed invariants persisting across failure
boundaries~\cite{borrill2026pratt}.

\subsection{The Four Industrial Anti-Patterns}

The practical consequence of FITO thinking manifests in four industrial
anti-patterns that pervade distributed systems
design~\cite{borrill2026pratt}:

\begin{enumerate}[label=(\roman*)]
\item \textbf{Timestamp re-importation.}  Attaching physical timestamps
  (NTP, GPS) and using them to define causal order.  This collapses
  structural independence information into a total order, destroying
  precisely what partial-order models reveal.  ``Last Writer Wins'' is
  the canonical example: conflicts resolved by timestamp comparison,
  treating clock readings as oracles of causality.

\item \textbf{Global ordering services.}  Systems like Apache Kafka,
  Google Spanner, and consensus-based logs provide a total order on
  events.  The log is treated as ``what happened'' rather than ``one
  consistent projection.''  Multiple pomsets---differing only in
  independent event order---are conflated into a single sequence,
  re-introducing the interleaving model that pomsets were designed
  to escape.

\item \textbf{Retry as forward-only convergence.}  Failure recovery via
  exponential backoff and retry assumes eventual convergence to a
  ``correct'' state.  Two transactions with no causal constraint are
  forced into total order by retry logic.  Independence is lost;
  information about failure causes is discarded.

\item \textbf{DAG re-sequentialization.}  DAG-based consensus protocols
  (IOTA Tangle, Hedera Hashgraph, Avalanche) use directed acyclic graph
  structure as intermediate scaffolding, then derive a linearization for
  state machine replication.  The DAG rhetorically embraces concurrency
  but operationally collapses to total order.
\end{enumerate}

Each anti-pattern reflects the same category mistake: treating FITO
artifacts as if they were inherent in the problem, rather than
recognizing them as consequences of the modelling assumption.

\section{Discussion}
\label{sec:discussion}

\subsection{What We Are Not Claiming}

We are not claiming that the impossibility theorems are wrong.  FLP is a
valid theorem; Halpern-Moses is a valid theorem; CAP is a valid theorem;
Herlihy's hierarchy is correct.  Every one of these results is
mathematically rigorous.

We are claiming that they are \emph{conditional}.  They are theorems
about FITO systems---systems whose communication primitive is the
unilateral message.  They do not apply to systems whose communication
primitive is the bilateral transaction, because the models in which they
are proved do not contain bilateral transactions.

This is precisely analogous to Spekkens' point about Bell's theorem:
Bell proves that FITO causality plus locality is incompatible with
quantum mechanics.  The standard conclusion is that locality fails.  The
alternative conclusion is that FITO fails.  Both are logically
valid.  The choice between them is a choice of which assumption to
drop---and that choice has consequences for the design space available.

\subsection{The Pattern Across Domains}

The category mistake pattern repeats:

\begin{enumerate}
\item An epistemic construct is mistaken for an ontic entity.
\item Engineering systems are built on the ontic reading.
\item Impossibility results and escalating compensation mechanisms
  reveal the strain.
\item The puzzles dissolve when the category is corrected.
\end{enumerate}

In quantum mechanics, the correction led from the measurement problem
to the epistemic interpretation.  In cislunar timekeeping, the
correction leads from synchronized time to relational
time~\cite{borrill2026cislunar}.  In distributed computing, the
correction leads from message-passing impossibilities to transactional
design.

\subsection{Kinematics and Dynamics Must Yield to Causal Structure}

Spekkens argued that the traditional distinction between kinematics (the
space of physical states) and dynamics (the law of evolution) cannot be
fundamental, because any change to kinematics can be compensated by a
corresponding change to dynamics without empirical
consequence~\cite{spekkens2012paradigm}.  What is physically significant
is the \emph{causal structure}---the pattern of which events can
influence which other events.

The same argument applies to distributed systems.  The traditional
distinction between ``state'' (the data at each node) and ``protocol''
(the rules for exchanging messages) is not fundamental.  One can always
trade a richer state space for a simpler protocol, or vice versa,
without changing the observable behaviour.  What is fundamental is the
\emph{interaction structure}---the pattern of which bilateral
transactions are possible between which endpoints.

This is the lesson of Herlihy's hierarchy, read through the Leibnizian
lens: the consensus number of a primitive is not a property of its
``state'' or its ``protocol'' but of its \emph{interaction structure}.
Read/write has consensus number~1 because it is unilateral.  Swap has
infinite consensus number because it is bilateral.  The physics is in
the interaction, not in the state.

\section{Conclusion}
\label{sec:conclusion}

Distributed computing has spent fifty years proving theorems about what
cannot be done and building increasingly complex mechanisms to
approximate what those theorems forbid.  FLP told us consensus was
impossible; we built Paxos and Raft to work around it.  The Two
Generals Problem told us common knowledge was unattainable; we built
protocols that settle for ``almost certainly.''  CAP told us we must
choose between consistency and availability; we built systems that
sacrifice one, then the other, depending on conditions.

We have argued that these impossibilities are not discoveries about
the physical world.  They are consequences of a category mistake:
treating FITO---the assumption that information flows forward in time
only, through unidirectional channels, from sender to passive
receiver---as a law of nature rather than a design choice.

The same conceptual move that dissolves quantum mysteries---recognizing
what is epistemic versus what is ontic, applying the Leibnizian
principle to eliminate surplus structure, taking interaction rather than
transmission as the primitive---dissolves the apparent impossibilities
in distributed computing.

The impossibility theorems are real.  They tell us exactly what cannot
be done with FITO primitives.  But they do not tell us what can be done
\emph{without} them.  The transactional alternative---bilateral
interactions where the primitive is the completed atomic exchange, not
the message in flight---opens a design space that the field has barely
begun to explore.

The category mistake is the same; the resolution is the same; only the
domain differs.

\section*{Acknowledgements}

\subsection*{The Mulligan Stew Gang}

The arguments in this paper were forged in conversation.  Since
mid-2023, a remarkable group of physicists, mathematicians, computer
scientists, network architects, and practicing engineers has met every
Friday morning at 7\,am Pacific Time---nearly without exception---to
interrogate the foundations of distributed computing, the nature of
time, and the category mistakes that connect the two.  This group, the
\emph{Mulligan Stew Gang}, grew out of the broader
\emph{It's About Time} community
(\url{https://itsabouttime.club}), a public forum that at its peak
drew over two thousand participants to live discussions on the nature
of time and causality across physics, computer science, neuroscience,
and philosophy.

Where \emph{It's About Time} cast a wide net, the Mulligan Stew
sessions became the crucible: a small, committed working group where
every claim in this paper was tested, challenged, extended, and
sometimes demolished before being rebuilt.  The FITO analysis, the
application of Spekkens' framework to distributed systems, the
reinterpretation of the impossibility theorems, the identification of
the four industrial anti-patterns, and the transactional alternative
all bear the marks of these Friday morning conversations.  Ideas that
survived the Mulligan Stew are stronger for it; ideas that did not
survive deserved their fate.

The members of this group---whose individual contributions span
quantum foundations, database theory, network protocol design,
causal inference, synchronization theory, entropy, and the philosophy
of time---prefer to remain uncredited by name.  Their intellectual
generosity, rigour, and willingness to challenge received wisdom over
more than one hundred and thirty consecutive weeks of sustained
collaboration have shaped every section of this paper.  The author's
debt to them is profound and gratefully acknowledged.

\subsection*{Edward A.\ Lee}

A special debt is owed to Edward A.\ Lee, Distinguished Professor
Emeritus of Electrical Engineering and Computer Sciences at the
University of California, Berkeley.  His book \emph{Plato and the Nerd:
The Creative Partnership of Humans and
Technology}~\cite{lee2017plato} provided the intellectual catalyst for
much of what followed.  Lee's rigorous insistence that the map is not
the territory---that models are human constructions, not mirrors of
nature---inspired both the Mulligan Stew Gang and the entire
D{\AE}D{\AE}LUS team to dig deeper, think harder, and eliminate
surplus ontology from the design of Open Atomic Ethernet.  His
forthcoming \emph{Deterministic
Concurrency}~\cite{lee2026deterministic}, a treatise on determinism
in concurrent systems now circulating in draft, sharpens the argument
further: that nondeterminism in computing is not an inevitable
consequence of concurrency but an artefact of modelling choices---a
claim that resonates directly with the thesis of this paper.  The
insights expressed here owe more to Lee's programme than a citation
can convey.

\subsection*{Use of AI Tools}

The theoretical framework presented in this paper---the application of
category mistake analysis, FITO assumptions, and the ontic/epistemic
distinction to the foundations of distributed computing---is the product
of more than twenty years of the author's independent research in
distributed computing, network architecture, and the foundations of
concurrency theory.  The core arguments, the identification of the
category mistake in distributed computing's impossibility results, and
the transactional alternative derive from the author's prior published
and unpublished work, including the Category Mistake monograph, the FITO
analysis series, the Pratt/Pomsets analysis, and the Open Atomic
Ethernet specification programme.

Large language models (Anthropic's Claude) were used as research
instruments during the preparation of this manuscript: for literature
search and verification, for testing the robustness of arguments
against counterexamples, and for drafting prose from the author's
detailed outlines and technical notes.  This usage is analogous to the
use of any computational research tool---a telescope extends the eye,
a calculator extends arithmetic, and a language model extends the
capacity to search, draft, and stress-test arguments at scale.  The tool
does not originate the ideas any more than a telescope originates the
stars.

All intellectual content, theoretical claims, original analysis, and
conclusions are the author's own.

\bigskip
\noindent\rule{\textwidth}{0.4pt}

\end{document}